\documentclass[useAMS,usenatbib]{mn2e}
\usepackage{epsfig,amssymb}

\title[An isolated compact elliptical galaxy]{Discovery of an isolated compact elliptical galaxy in the field}
\author[Huxor et al. ]{A. P. Huxor$^{1}$ \thanks{email: avon@ari.uni-heidelberg.de }, S. Phillipps$^{2}$  and J. Price$^{2}$  \\\
$^{1}$Astronomisches Rechen-Institut, Zentrum f\"{u}r Astronomie der Universit\"{a}t Heidelberg, M\"{o}nchstra{\ss}e 12 - 14, \\
69120 Heidelberg, Germany.\\
$^{2}$University of Bristol, H. H. Wills Physics Laboratory, Tyndall Avenue, Bristol, BS8 1TL, U.K. \\
}
\begin{document}

\date{}

\pagerange{\pageref{firstpage}--\pageref{lastpage}} \pubyear{2002}

\maketitle

\label{firstpage}

\begin{abstract}
We present the discovery of an isolated compact elliptical (cE) galaxy, found during
a search of SDSS DR7 for cEs, and for which we obtained WHT/ACAM imaging.
It is $\sim$900 kpc distant from its nearest neighbour, has an effective r-band radius of $\sim$ 500 pc and a B-band mean surface brightness within its effective radius of 19.75 mag arcsec$^{-1}$. Serendipitous deep SuprimeCam imaging shows that there is no underlying disk. Its isolated position suggests that there is an alternative channel to the stripping scenario for the formation of compact ellipticals.  We also report analysis of recent deeper imaging of the previous candidate free-flying cE, which shows that it is,  in fact, a normal dwarf elliptical (dE). Hence  the new cE reported here is the first confirmed isolated compact elliptical to be found in the field.

\end{abstract}


\section{Introduction}

Compact elliptical (cE) galaxies are  a comparatively
  rare class,  possessing very small effective radii and  
high central surface brightnesses \citep{Faber73}. These are defined as galaxies that  sit at the low-mass of the elliptical sequence in scaling relations and the prototype is the Local Group dwarf galaxy M32.

\begin{table*}
 \centering
  \caption{
 Basic Properties
  }
 % \hline
  \begin{tabular}{@{}llllll@{}}
ID					& Alt Name 		&  RA 		&  Dec   	  		& D$_{L}$ (Mpc) 	& scale (kpc arcsec$^{-1}$)			\\
 \hline 
J094729.24+141245.3	& cE0 			& 09 47 29.24 	& +14 12 45.3		& 85.5  		& 0.399				\\
NGC6868-f2-0881		& 6868/18			& 20 09 01.12   & --48 19 15.9		& 40.8		& 0.194				\\
\hline
\end{tabular}
\end{table*}\label{Tab:basic_data}

Two views on compact ellipticals are found in the literature. 
As many known cEs  occur close to massive galaxies in groups or clusters, many believe they are the result of the tidal stripping and truncation that results from interactions with their giant neighbours, and computer simulations support the possibility  \citep{Faber73,Bekkietal01, Choietal02, Priceetal09}. Alternatively, they have been considered as either the natural extension of the class of elliptical galaxies to lower luminosities and smaller sizes \citep{WirthGallagher84,Kormendyetal09}. \citet{KormendyBender12} argue against the stripping scenario as: not all cEs are companions of massive galaxies; many dSph galaxies are companions to massive galaxies yet are not truncated; bulges also sit at the compact end of the scaling relations for elliptical galaxies but are not  truncated; and that  unlike globular clusters, the exemplars for ideas of tidal truncation, the dark matter halo of a dwarf galaxy will quickly lead to a merger with its massive host.

Of course, it could be that both formation scenarios  lead to the observed class of  cE galaxies. Although we recently  reported two cEs, which show irrevocable evidence of  formation resulting from the ongoing tidal stripping of more massive progenitors -- with the tidal tails clearly visible \citep{Huxoretal11},  we also commented that stripping may not be the only explanation. In this paper we present a compact elliptical galaxy found in the field, whose isolation shows that it is not stripped, but a classical -- if compact -- elliptical galaxy.

One excellent way to identify the cEs that are examples of the low-mass end of the classical ellipticals is to identify  and characterise cEs that are truly isolated and cannot in any way have been affected by a more massive host.  Hence, we have undertaken a search for cE candidates in SDSS DR7 (Huxor et al. in prep), which, being a wide-field survey, contains considerable data in the field. However, the SDSS imaging is typically not sufficiently deep to exclude the presence of a faint underlying disk at the distances for which our candidates are found. The isolated cE
candidates only have a small chance of being found in deep archival
data -- such imaging tends largely to target denser environments,
rather than the field. However,  one candidate isolated cE (hereafter called cE0)  was found in archival  Subaru Prime-Cam imaging, allowing it to be studied in detail and reported here.

\section{Data and Analysis}

\subsection{SDSS}

Our initial search for cE candidates was taken from SDSS DR7. The main $\it{Legacy}$ survey covers $\sim$ 8500 square degrees of sky, with spectroscopy of many complete samples of galaxies \citep{Abazajianetal09}. Using the SQL query language on the $\it{galaxy}$ sub-sample, a sample of compact galaxies was assembled. Further details are given in \citet{Huxoretal11}.

The SDSS catalogue data gives an effective radius of 1.25 arcsec for cE0, which corresponds to  520 pc, at our adopted distance of 85 Mpc (see below) -- sufficiently compact to be of interest. It also has a value for the SDSS catalog parameter \emph{fracDeV} of 1, indicating that it   does not posses a value for S\'ersic index {\it n}, which would indicate it is a distant spiral, for example.

This radius is at the limit of measurability by the SDSS pipeline, hence we re-determine the value from the SDSS imaging, using GALFIT  \citep{Pengetal02}. GALFIT is a 2D galaxy fitting code which creates a convolution of a  S\'ersic model with the point spread function of the source image (and so should be understood to be seeing corrected). The best fit is determined by comparing this convolved image with the science image of the galaxy, minimising the $\chi^{2}$ of the fit.  The SDSS pipeline only fits to fixed S\'ersic n values of 1 (exponential) or 4 (de Vaucouleurs), while GALFIT allows a free fit on the n parameter (S\'ersic n is known to vary, even among ellpticals -- see \citealt{Caonetal93}), and the best fit model gave an effective radius of $\sim$ 500 pc, with a S\'ersic n of 4.35. The S\'ersic n value confirmed that we were seeing an elliptical-like galaxy, and the effective radius makes it comparable to many cEs in the literature.

cE0 has a SDSS redshift of 0.0198, which using a 5 year WMAP cosmology  \footnote{Throughout  we adopt a WMAP 5 year cosmology, H$_{0}$= 70.5 , $\Omega_{matter}$=0.27 and $\Omega_{\Lambda}$=0.73, \citep{Hinshawetal09}, and use the PYTHON code version of the \citet{Wrightetal06} cosmology calculator.} gives 
 a luminosity distance of 85.5 Mpc and an angular diameter distance of 82.2 Mpc, equivalent to a scale of 0.399 kpc arcsec$^{-1}$ (Table \ref{Tab:basic_data}). The galaxy has an SDSS catalogue (g-r)$_{0}$ colour of 0.74 and an r-band magnitude of 15.96, giving an extinction corrected absolute magnitude in the r band of --18.78.

The redshift is equivalent to cz=5937 kms$^{-1}$.  Allowing for a very conservative group velocity dispersion of up to 300 kms$^{-1}$ \citep{Tagoetal08}, to identify potential members of a group within which cE0 might sit,  NED finds  one galaxy (CGCG 063-062) at a projected distance of 35 arcmins ($\sim$840 kpc using our adopted distance modulus), from cE0, with a r-band magnitude of 13.88, and another (UGC 05258) 39 arcmins ($\sim$930 kpc) from cE0. The actual physical distances between the galaxies are, of course, likely to be much greater. These distances are very different from those we found \citep{Huxoretal11} for the projected distances for the two cEs that are clearly being stripped by a massive neighbour (15 to 20 kpc). It is difficult to imagine a scenario in which cE0 could be so close as to be stripped by one of the more massive galaxies, and then move to its current distance from them.

\subsection{WHT-ACAM}

The seeing for the SDSS imaging was $\sim$1.1 arcsec, which at the distance of cE0 corresponds to $\sim$450 pc, similar to the
effective radius of the galaxy, so we should check that our measurement is not limited by the SDSS seeing.

\begin{figure}
\centering
 \includegraphics[angle=0,width=60mm]{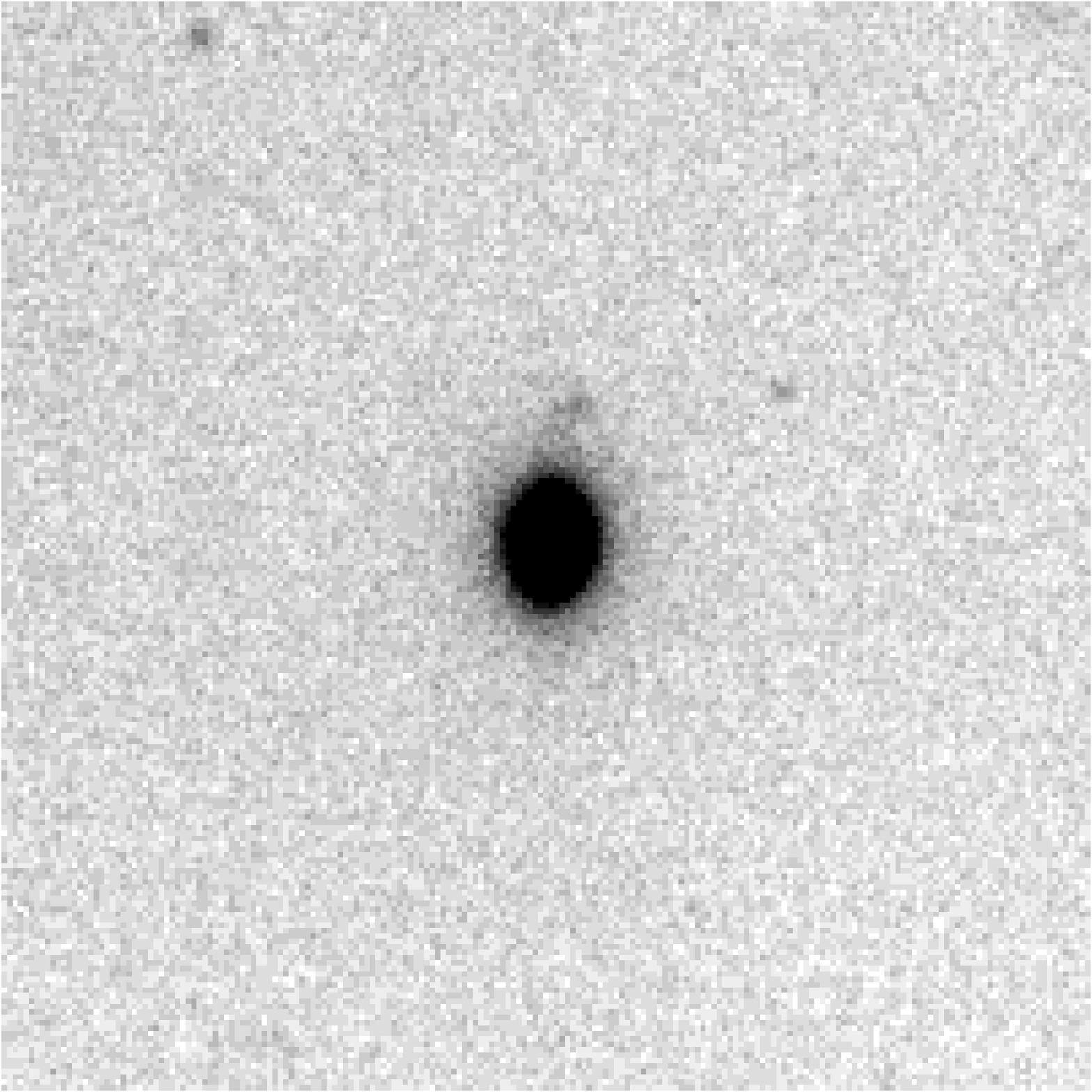}
 \vspace{1pt}
 \caption{WHT ACAM r-band of mage of cE0. 20 $\times$ 20 kpc. North up.  East is left.}\label{Fi:ACAM}
\end{figure}

\begin{figure}
\centering
 \includegraphics[angle=0,width=80mm]{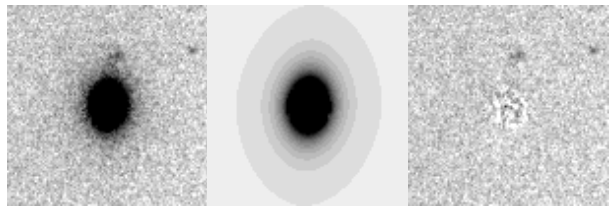}
 \vspace{1pt}
 \caption{GALFIT fit to the WHT ACAM Image of cE0, 10 $\times$ 10 kpc.  Same orientation as Figure \ref{Fi:ACAM} }\label{Fi:galfit}
\end{figure}

 However, the archival Subaru data could not be used to determine the structural parameters
as the central region of cE0  was saturated.  Thus we sought new imaging data in service mode with a 
requirement for  good seeing from the new ACAM camera on the William Herschel Telescope (WHT). The data was reduced in the standard manner using IRAF\footnote{IRAF is distributed by the National Optical Astronomy Observatory, which is operated by the Association of Universities for Research in Astronomy (AURA) under cooperative agreement with the National Science Foundation.} using flats and biases taken on the same night.
The seeing (FWHM) was 0.85 arcsec in the r-band , i.e. a seeing half-radius of $\sim$ 170 pc. We are confident that GALFIT will return reliable parameter values; \citet{Trujilloetal06} for example, found --  from simulated data -- that they could estimate accurate parameters when the effective radius of the galaxy was comparable in size to the seeing half-radius. 

We  derive parameters using this ACAM r-band imaging, using a stacked image of the four individual exposures that had been taken. 
The stacked image was analysed  with GALFIT to obtain structural parameters, and obtain revised photometry. In each case, the PSF is derived from stars within the image field. 
Although the SDSS provides good photometry for the galaxies, we also use the model fits to determine revised total model magnitudes, to be consistent with the specific models fit by GALFIT. Calibration was achieved by PSF fitting of bright and  isolated stars, with known SDSS magnitudes, to set the appropriate photometric zeropoint. The results are given in table 2

\begin{table}
 \centering
  \caption{
 Photometric Properties
  }
 % \hline
  \begin{tabular}{@{}lllll@{}}
Name 		& $M_{B0}$	& $R_{e}$			& S\'ersic $n$		& $<\mu_{e,B}>$			\\
   			&			& pc				&  				& mags arcsec$^{2}$  		 \\
 \hline 
 cE0			&-17.72$\pm0.03$		& 499$\pm10$ 		& 4.3$\pm0.1$ 		& 19.75$\pm0.03$			\\
6868/18		& -16.11$\pm0.12$		&  764$\pm6$		& 1.6$\pm0.1$			& 21.96$\pm0.12$					\\
\hline
\end{tabular}
\end{table}\label{Tab:derived_data}

A revised r-band magnitude for cE0 was obtained from the ACAM data. The colour from the SDSS was then used to convert this to B-band values using the colour transformation equations in  \citet{Jesteretal05}, to allow for comparison with the literature, and the result is given in Table \ref{Tab:basic_data}, and plotted in figure \ref{Fi:plots}.

 The errors on the parameters returned by GALFIT have been found to underestimate the actual uncertainty by a large factor \citep{Haussleretal07}. Hence, the errors for the GALFIT parameters were estimated by applying GALFIT to the four individual images, then the six combinations of any two images, and finally  the four combinations of any three images. The standard deviations found for the effective radii, magnitudes and S\'ersic n were then extrapolated to the final image of four stacked science exposures. For the magnitudes, the uncertainties are mainly due to those from the colour transformation.

From the plot shown in figure \ref{Fi:plots} we can see that for its absolute magnitude, cE0 has a mean surface brightness and effective radius that
place it in the region occupied by  the known compact ellipticals.

\subsection{Archival Suprime-Cam Imaging}

\begin{figure} 
\begin{center}$
\begin{array}{cc}
 \includegraphics[angle=0,width=39mm]{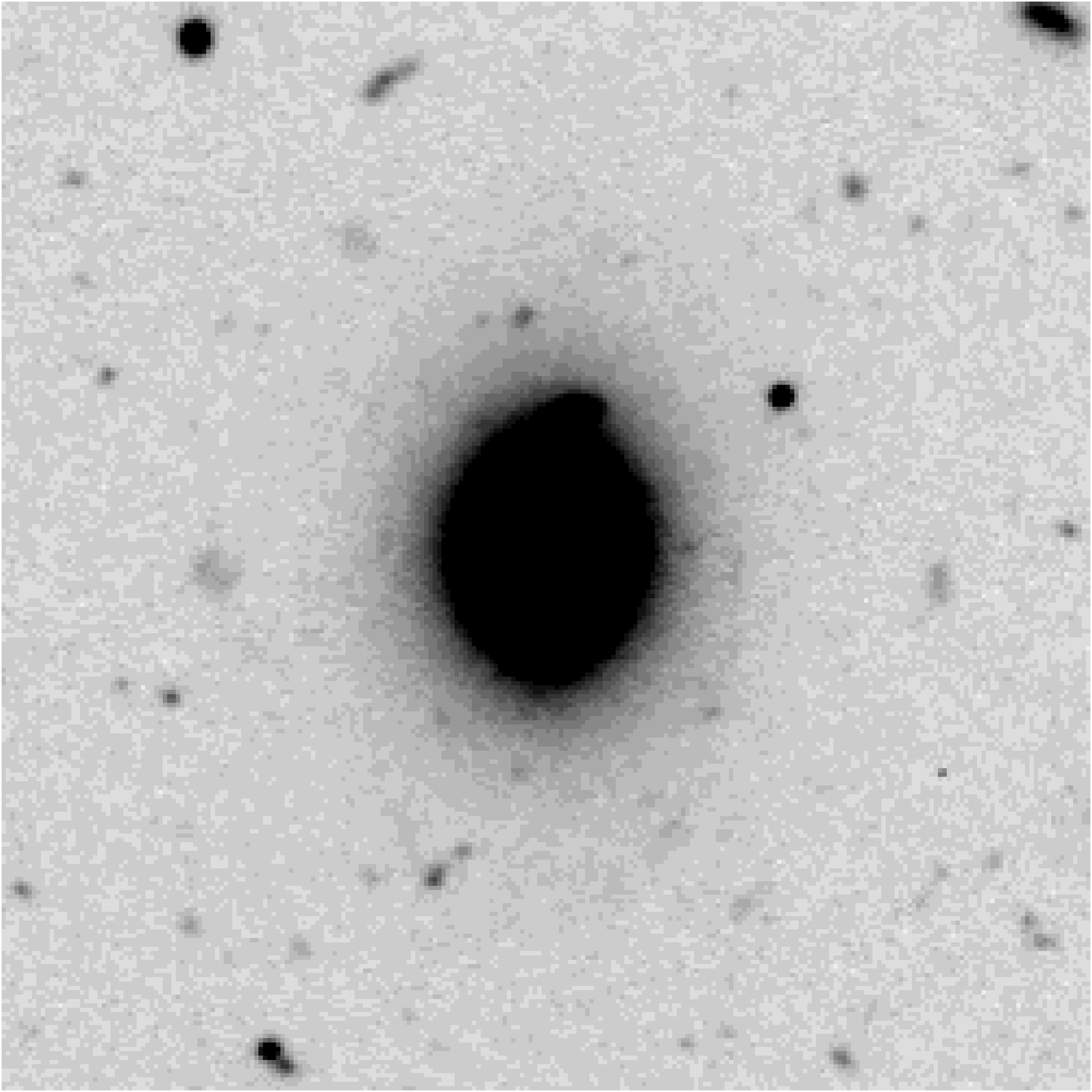} &
 \includegraphics[angle=0,width=39mm]{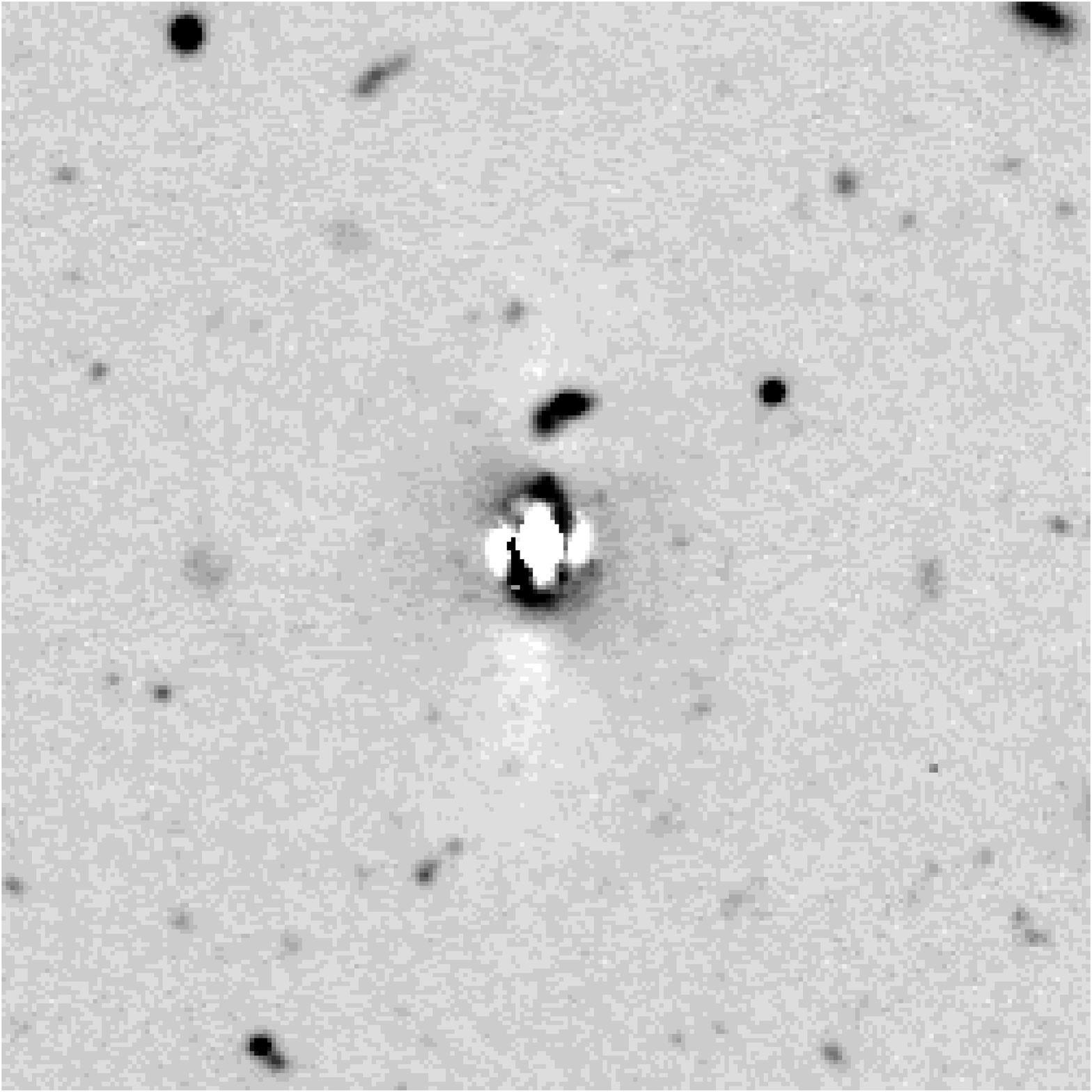}
\end{array}$
\end{center}
\caption{Left: Subaru archival image of cE0. 20 $\times$ 20 kpc. Same orientation as Figure \ref{Fi:ACAM} .
 This image comprises a total of 750 seconds, and yet there is no evidence of a disk. Right:Residual of Subaru image.  The centre of the galaxy is saturated and so masked out, leading to the residuals seen. The structure to the north appears to be background galaxies.} \label{Fi:Subaru_images}
\end{figure}

As noted above, Subaru/Suprime-Cam archival imaging of cE0 is available. The Suprime-Cam \citep{Miyazakietal02} data was reduced with SDFRED1\citep{Yagietal02,Ouchietal04}, the appropriate tool for observations taken on the relevant dates. The Subaru observations of cE0 were observed in one passband\footnote{The filter used is W-J-VR, which is a ``user" filter and not a standard Subaru filter. It has a range from $\sim$5000 to 7000 $\AA$, and encompasses both the Johnson V and Cousins R passbands.}, so no colour information is available.
  The depth of the exposures are such  that the centre of cE0 is saturated. Thus we are unable to derived structural parameters from this data. 
 But it is clear that there is no low-surface brightness disk visible (see Figure \ref{Fi:Subaru_images}). 
 
The Subaru image was masked to remove the saturated inner region of cE0, and GALFIT was applied to this masked image (see Figure \ref{Fi:Subaru_images}), with the structural parameters fixed to those derived from the ACAM image. The residuals near to the centre are most likely an artifact from the large masked region. This image does, however, show no obvious outer structure. The object to the north of cE0 is now also seen almost certainly to be two background galaxies, possibly associated with the others in the field, some of which are visible in the figures.

\subsection{Spectroscopy}

We analyse the SDSS spectra for cE0 to determine ages and key stellar population indicators and their errors as described in  \citet{Priceetal09,Priceetal11}, but we summarize briefly here. The models of \citet{Schiavon07} and the EZ-Ages code \citep{GravesSchiavon08} are used to determine the age, metallicity and key abundance indicator [Mg/Fe] of the galaxy. They are the values of these parameters for a SSP (single stellar population) which most closely reproduce the data, while the value for the velocity dispersion is taken  from SDSS (table  \ref{Tab:spectral}). 

Although similar to stripped cEs in many regards, cE0 also shows differences in its stellar populations. cE0 appears both older, and of lower metallicity, than the cEs known to be stripped (and reported in \citealt{Huxoretal11}).

\begin{table} 
 \centering
  \caption[Spectroscopic Properties of cE0]{
Spectroscopic Properties of cE0.
  }
 % \hline
  \begin{tabular}{@{}lllllllll@{}}
     ID 	& Age (Gyr)	 		&  [Fe/H]			& [Mg/Fe] 			 & $\sigma$ (kms$^{-1}$)  	 \\
 \hline 
  cE0	 	& 9.21$^{+6.3}_{-3.87}$	& -0.19$\pm0.16$	& 0.09$\pm0.14$ 	&  105$\pm6$   				 \\
\hline 
\end{tabular}\label{Tab:spectral}
\end{table}

\section{Previous candidate free-flying compact elliptical}

\begin{figure}
\centering
 \includegraphics[angle=0,width=60mm]{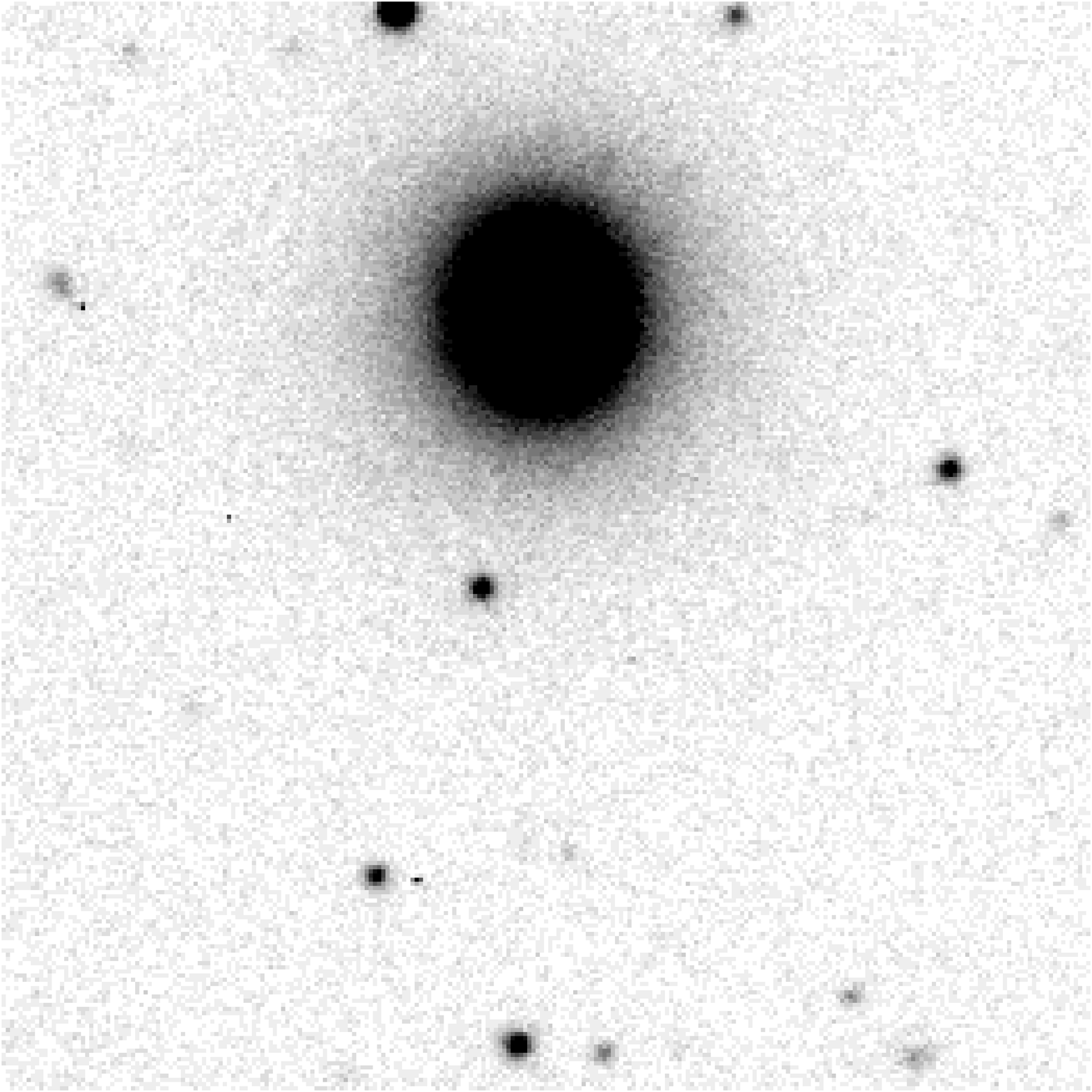}
 \vspace{1pt}
 \caption{R-band FORS1 image of NGC 6868/18, 10 $\times$ 10 kpc. the galaxy lies near the edge of the CCD as so is not centred in the image. North is up, and East is left.}\label{Fi:6868ce}
\end{figure}

\begin{figure}
\centering
 \includegraphics[angle=0,width=80mm]{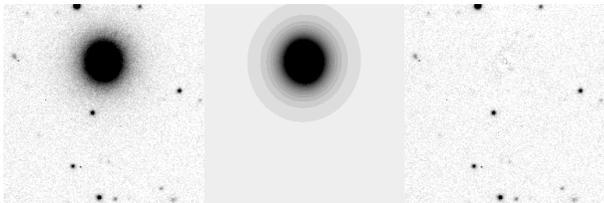}
 \vspace{1pt}
 \caption{GALFIT fit to the FORS1 image of NGC 6868$\_$18. The images have the same size and orientation as Fig. \ref{Fi:6868ce}. }\label{Fi:galfit6868ce}
\end{figure}

There is one precedent for an isolated, so-called ``free-flying"  compact elliptical in the literature. \citet{WirthGallagher84} describe a  cE in the NGC 6868 group\footnote{That paper gives the group as NGC 6861 but from \citet{Wirth83}, it is clear that NGC 6868 is intended, and this is confirmed by Figure 3 of  \citet{WirthGallagher84} .}, but this does not
seem to be have been subsequently followed up. A plate image of n6868/18 is seen in  \citet{Wirth83}, and when compared with the DSS images we get coordinates for this galaxy as listed in Table \ref{Tab:derived_data}.

\begin{figure*} 
\begin{center}$
\begin{array}{cc}
\includegraphics[angle=90,width=87mm]{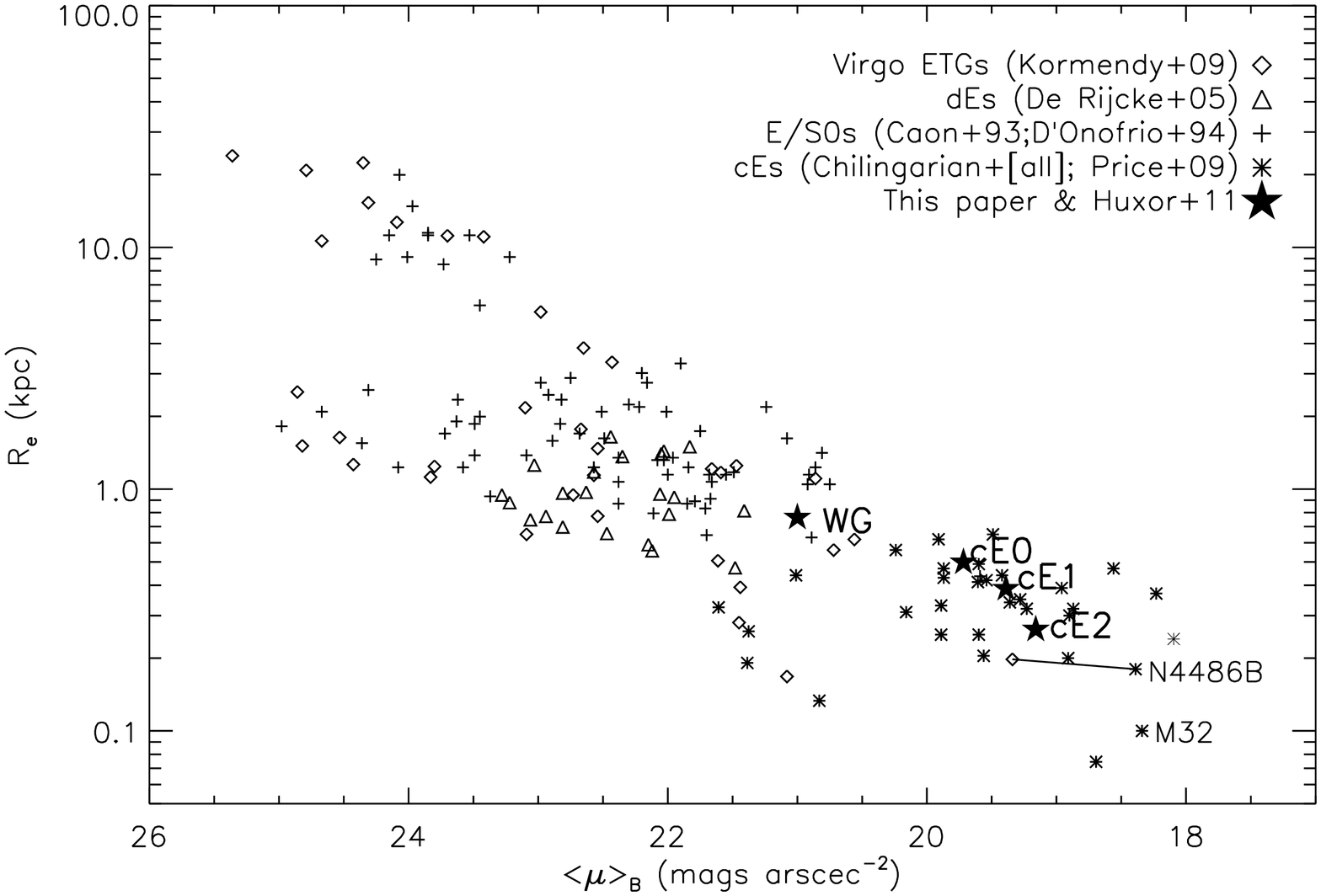} &
\includegraphics[angle=90,width=87mm]{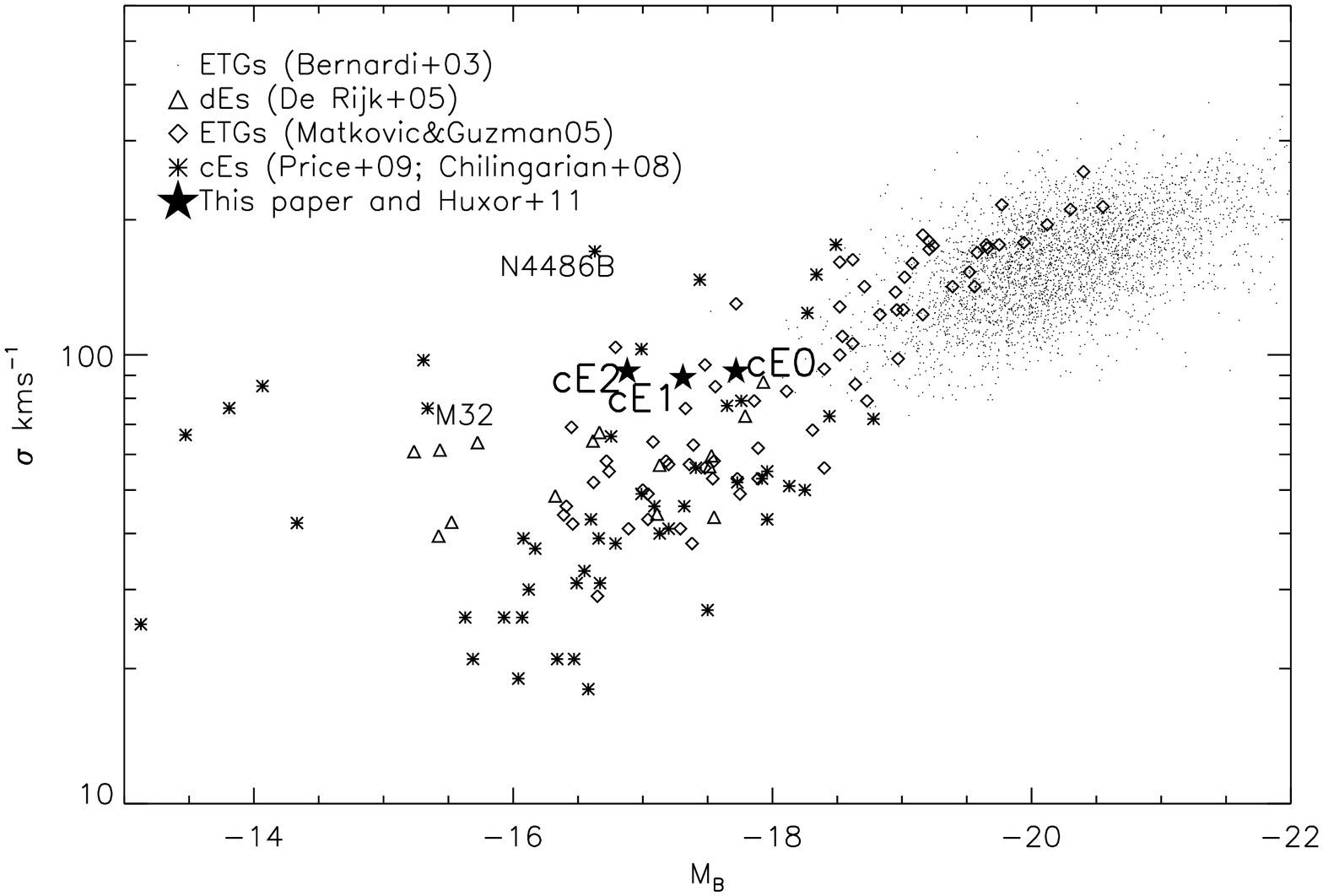}
\end{array}$
\end{center}
\caption{Relation of cE0 (labelled in plot), and the \citet{WirthGallagher84} dwarf galaxy (WG), to other hot stellar systems. These include the  two cEs (cE1 and cE2) from \citet{Huxoretal11}. Also shown are cEs (asterisks) \citep{ChilingarianBergond10, Chilingarianetal09, Chilingarianetal07,Priceetal09} and other ETGs (diamonds and triangles )  \citep{Kormendyetal09, deRijckeetal05, Caonetal93, DOnofrioetal94}.  Different datasets are shown by different symbols, and all are converted to our chosen cosmology (5 year WMAP).  
Left: Plot of stellar systems, including ETGs and known cEs in surface brightness vs effective radius.  Colours, allowing the determination of B-band magnitudes for the \citet{Kormendyetal09} data were obtained from the GOLDMine database (goldmine.mib.infn.it), where available. For the four galaxies where this data not available the (B-V) is estimated from a linear fit of V-band magnitude and (B-V) colour derived from the rest of that sample. We exclude the galaxies classified as S0 in the \citet{Kormendyetal09} sample. Values for M32 and NGC 4486B from \citet{Chilingarianetal07} are labelled, with a black line connecting the data points for NGC 4486B from \citet{Chilingarianetal07} and \citet{Kormendyetal09}.
		Right: The Faber-Jackson relation. We also include data from \citet{Bernardietal03} for a redshift $< 0.1$ (using relations in \citealt{Jesteretal05} to convert colours), \citet{MatkovicGuzman05} (where S/N $>$ 15), and 
 \citet{Chilingarianetal08} (for early type galaxies in their sample only).} \label{Fi:plots}
\end{figure*}

The NGC 6868 candidate cE was found in archival VLT/FORS1 \citep{Appenzelleretal98}  imaging. This data was reduced in the usual manner with IRAF  using biases and flat fields taken on the same night as the science data. There were two R-band images with the galaxy sufficiently within the frame to allow for photometry of the whole galaxy. These were co-added to give a total exposure time of 180 secs.  This image was analysed with GALFIT in the same manner as cE0, with the uncertainty for the effective radius estimated by taking the multiplier from the GALFIT reported uncertainty to that we found by our method described above. For the value of n, GALFIT reported an 
optimistic uncertainty of 0.00 for a single S\'ersic fit. More realistically, a two 
S\'ersic fit in which one component was much more luminous than the other gave an 
uncertainty which, when scaled up as described for cE0, gave a similar error to that for 
cE0 itself.

This galaxy has a  V-band  magnitude of 16.07 \citep{Carrascoetal06}. We can convert this to a B-band magnitude using the colour  (B-V) of 0.70 ($\sigma$=0.12)   given in  \citet{Karicketal03} for  the dwarf ellipticals of the Fornax Cluster.
Using this value gives a mean surface brightness of $\sim$ 22.0 mag per arcsec$^{2}$ in the B-band. For the magnitudes, the errors are driven by the uncertainty in the colour of  \citet{Karicketal03}.

The results are presented in table 1, and plotted in figure \ref{Fi:plots}.
We see that with the new data, the
object no longer appears particularly compact. The surface brightness and size place it amongst the dwarf ellipticals. It also has a S\'ersic $\emph{n}$ of 1.6,  consistent with this class of galaxy. 

\section{Conclusions}

We present an isolated compact elliptical galaxy  found in the field -- rather than group or cluster environment.  The nearest galaxy to cE0 (CGCG 063-062) is at $\sim$840 kpc projected distance, and hence almost certainly more distant still. This makes the stripping scenario very unlikely, suggesting that cE0 formed in the manner suggested by \citet{Kormendyetal09}. 

When we consider its location in both the Kormendy and Faber-Jackson relations (see Figure \ref{Fi:plots}) we see that the new cE is near the relation defined by the published sample of cEs. It is very close in these plots to cE1 and cE2, the two cEs in which we see their formation by stripping \citep{Huxoretal11}.

However, despite the similarity in structural parameters, cE0 does  differ from cE1 and cE2, in that it is both older and more metal-poor. Could it be that some of the ancient cEs found by \citet{Chilingarianetal09} are not stripped disk galaxies, but low-luminosity, classical ellipticals that were accreted into the galaxy clusters in which they were subsequently found? That is, the apparent age may not be due to their being formed by stripping at an early epoch, but due to them being classical old, but compact, ellipticals \citep{Kormendyetal09}.

\citet{WirthGallagher84} made the comment that cEs would not be easy to detect, and that their ``predilection" for being found near massive galaxies is a result of the latter being the usual targets for large-scale imaging. This newly discovered galaxy, cE0, seems to prove the case, and argues for the need to investigate the field to discover further examples. If found, their properties, as a population, will help throw light on the formation of the faint end of the elliptical class of galaxies.

\section*{Acknowledgments}

APH would like to acknowledge the generosity of the Leverhulme Trust, during the course of whose grant this work was undertaken. This work was also partly supported by Sonderforschungsbereich SFB 881 "The Milky Way System" (subproject A2) of the German Research Foundation (DFG).

The WHT and its service programme are operated on the island of La Palma by the Isaac Newton Group in the Spanish Observatorio del Roque de los Muchachos of the Instituto de Astrof'sica de Canarias. This work is based in part on data collected at Subaru Telescope, which is operated by the National Astronomical Observatory of Japan. 
Funding for the SDSS and SDSS-II has been provided by the Alfred P. Sloan Foundation, the Participating Institutions, the National Science Foundation, the U.S. Department of Energy, the National Aeronautics and Space Administration, the Japanese Monbukagakusho, the Max Planck Society, and the Higher Education Funding Council for England. The SDSS Web Site is http://www.sdss.org/. 

Finally, we thank the referee, Prof. John Kormendy, whose comments greatly improved this paper.

\bsp

\label{lastpage}

\end{document}